# Adaptive FPGA NoC-based Architecture for Multispectral Image Correlation

*Linlin Zhang[1], Anne-Claire Legrand[1], Virginie Fresse[2], Viktor Fischer[2];*
*LIGIV[1], LaHC CNRS 5516[2]; Saint Etienne, France*

## Abstract

*An adaptive FPGA architecture based on the NoC (Network-on-Chip) approach is used for the multispectral image correlation. This architecture must contain several distance algorithms depending on the characteristics of spectral images and the precision of the authentication. The analysis of distance algorithms is required which bases on the algorithmic complexity, result precision, execution time and the adaptability of the implementation. This paper presents the comparison of these distance computation algorithms on one spectral database. The result of a RGB algorithm implementation was discussed.*

## Introduction

The multispectral images are acquired optically in more than one spectral or wavelength interval (see Fig.1). Each individual image is usually of the same physical area and scale but of a different spectral band. This type of imaging is particularly critical for high-end colour reproduction, multi-ink printing and hyper-spectral satellite observation for surface features identification [1] [2].

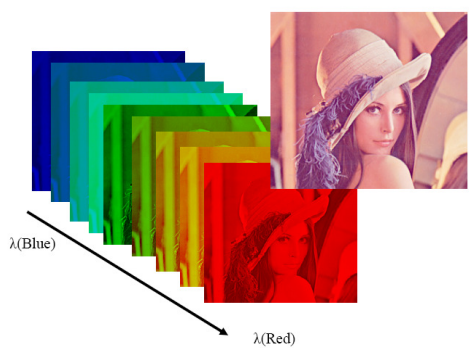

*Figure 1. Multispectral image example*

Museum and archive applications represent an important field of multispectral technology utilization [3]. For the image correlation of this kind of application, certain algorithms require high precision operations which imply large amount of floating-point data and complex functions (e.g. square root or other nonlinear functions). The computing depends on data size and clock frequency of the processor. In a single-processor system (e.g. PC of simple cores), the data are processed serially. NoC architecture on a FPGA (Field Programmable Gate Array) may become a choice to take advantage from the tasks parallelism to reduce the execution time.

This paper presents the implementation of the distance algorithms in the processing modules. The rest of the paper is organized as follows: Section 2 presents the adaptive GALS-based (Globally Asynchronous Locally Synchronous) architecture. Section 3 describes the multispectral image correlation authentication process which is implemented in the architecture. Section 4 presents an analysis of different colour spaces and a comparison of some multispectral distance algorithms. Section 5 presents a part of the RGB implementation results and discussion. Conclusions and perspectives are given in Section 6.

## The Adaptive GALS-based Architecture for the Image Analysis Algorithm

The proposed architecture was designed especially for the image analysis algorithms. The first image algorithm implemented in the architecture is the PIV processing (Particle Image Velocimetry) [4]. The advantage of the NoC architecture is that the whole system is adaptive. Certain modifications on the module level can adapt the architecture to the new multispectral applications.

### *Structure of the GALS-based Architecture*

There are four types of modules in this image analysis algorithm architecture (see Fig.2):
- Acquisition module: connects the acquisition system (multispectral camera, multispectral filter) with multispectral imaging system,
- Storage module: contains data memory
- Control module: controls the entire architecture,
- Processing modules: runs the calculations using the selected multispectral distance algorithm. Every module integrates one distance algorithm.

To implement several colour distance algorithms, the number of the processing modules is "theoretically" unlimited.

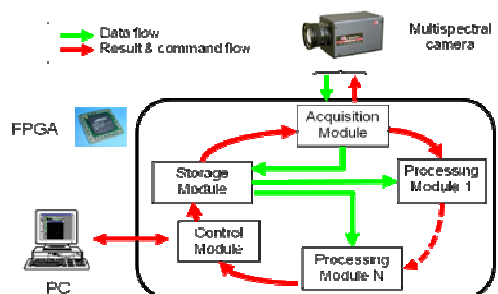

*Figure 2. The adaptive architecture for image analysis algorithms*

The lines as a circle in Fig.2 present the data flow (image) which has a high bandwidth. The ones inside the architecture present a mix result and command flow which has a small bandwidth.

The adaptive architecture is designed on the foundation of reusable IP (Intellectual Property) blocks and a pre-defined interface in GALS approach [5] [6]. The module structure is shown in Fig.3. One module contains several units (basic functions such as decode, interface, etc.). One unit contains several blocks (basic functions such as state machine, FIFO, etc.) which is the basis linear effort property. More details about the structure (modules and communication protocols) are given in [4].

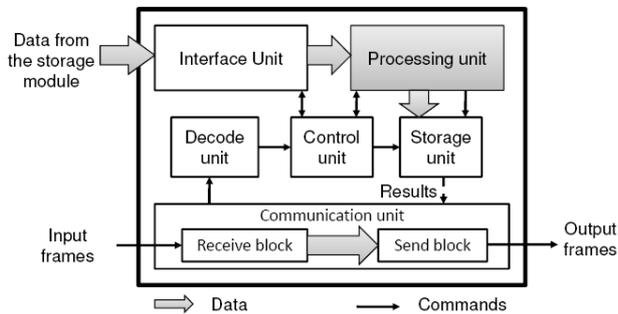

*Figure 3.* The structure inside the processing module of the adaptive architecture: multispectral algorithms are implemented in the processing units.

The architecture is designed by means of a C-based hardware description language [7] [8] and hardware description language. According to our requirement, Handel-C and VHDL are chosen. More electronic design details are presented in [9].

## FPGA Target Technology

The Xilinx Virtex-4 FPGA is the target technology [10]. Virtex-4 FPGA is a device containing user-programmable gate arrays with various configurable elements and embedded cores optimized for high-density and high-performance system designs. Virtex-4 devices contain the following components:
- Slice: contains two functions generator (F&G) and two storage elements. The function generators F&G are configurable as 4-input look-up tables (LUTs). The two storage elements are edge-triggered D-type flip-flops (Flip-Flop).
- Block RAM modules
- Cascadable embedded Digital Signal Processing slices (DSP48) with 18-bit * 18-bit dedicated multipliers, integrated Adder, and 48-bit accumulator.
- Digital Clock Manager (DCM), presented as 32 GCLKs (Global Clocks).

On a Virtex-4 XC4VLX15 FPGA, 6144 slices and 32 DSP48 are available. These resources are used to construct the architecture.

Among the mathematical operations, divide, square root, and accumulate are critical operations in many high performance signal processing applications. Floating point calculation was considered as a short-back of FPGA implementation. Floating-point division and square root are hard to implement due to the complexity of the algorithms. Early implementations of floating point operations on FPGAs used non-standard formats, largely because implementing IEEE compliant single precision add and multiply was impractical [11] [12]. Floating-point accumulators and multiply-accumulators have been previously discussed and implemented [13] [14]. Both the division [15] and square root [16] algorithms are based on lookup table and Taylor series expansion, and use a combination of small table lookup and small multipliers to obtain the first few terms of the Taylor series. Previous work proved that converting from floating-point to fixed-point or integer would greatly benefit the FPGA performance [17]. These algorithms presented in the recent floating-point (fix-point form) library [18] are particularly well-suited for implementation on a FPGA with embedded RAM and embedded multipliers such as Altera Stratix and Xilinx Virtex family devices.

## Multispectral Image Correlation

The aim of the multispectral image correlation is to compare two spectral images:
- Original image (OI): its spectra have been saved in the Storage Module as the reference data.
- Compared images (CI): its spectra are acquired by multispectral camera which is connected to the acquisition module.

For the art authentication process, OI is the information of the true printing (Original Image), and the CI are the others "similar" printings (Compared Image). With the comparison process of the authentication in the FiG.4, we need to find the true one by using multispectral imaging technique.

*Figure 4.* Basic comparison process of the authentication

The authentication principle in Fig.4, is based on the comparison of the two images in certain projection spaces (spectral space, RGB space, XYZ space, L*a*b* space, etc.). The colour space dimensions can vary from 3 to N. N presents the total multispectral imaging wavelength depending on the multispectral acquisition system (400 maximum in our condition).The comparison process follows four steps below:
1) Colour projection: with the chosen colour space references and the number of wavelength, original spectra data of OI/CI are transformed to certain colour image data
2) Multispectral distance calculation: with the chosen algorithms, image distance result $R_1$ is obtained.
3) Multispectral authentication: compare the result $R_1$ and the precision $P_1$. If the result is good enough for the required precision, the process can be ended.
4) Otherwise, the spectral number can be increased to have a higher precision calculation.

The first step (colour projection) is not always required for certain distance algorithms. The input data of the multispectral camera are 8-bit integers which present spectra values. If the distance calculation is based on spectra, the process can begin from the second step directly, such as RMS algorithm. But colour projections have their own advantages. The interest of three-dimensional colour representations is to help the observer to analyse the image. Some colour clusters (or regions of a given image) could be globally or separately better discriminated (or segmented) in one colour space. In the RGB processing module, one unit for the RGB projection is implemented.

## Colour Spaces and Distance Algorithms Analysis

As this comparison process is implemented in the processing module on VHDL/Handel-C in FPGA, an analysis of different colour projections and distance algorithms via electronic design is done.

### Colour Space Analysis

A colour model is an abstract mathematical model describing the way colours can be represented as tuples of numbers, typically as three or four values or colour components (e.g. RGB and CMYK). Adding a certain mapping function between the colour model and a certain reference colour space results in a definite "footprint" within the reference colour space. Some of the properties (linearity of transformation, stability of calculations, perceptual uniformity) of the various transformations from RGB to other colour spaces are summarized in Table 1. Note that, this analysis of colour spaces is just based on their own mathematical properties.

**Table 1: Transformation from RGB to other colour spaces**

| Colour space | Linearity | Stability | Uniform |
|---|---|---|---|
| rgb | No | No | No |
| XYZ | Yes | Yes | No |
| xyz | No | No | No |
| L*a*b*/L*u*v* | No | Yes | Yes |
| YIQ/YUV/$YC_bC_r$ | Yes | Yes | No |
| $AC_1C_2$ | Yes | Yes | No |
| $X_1X_2X_3/I_1I_2I_3$ | Yes | Yes | No |
| IHS, HSV | No | No | No |
| Munsel | No | Yes | Yes |

From Table 1, we chose 3 possible general spaces: RGB, XYZ and L*a*b* (or L*u*v*). Table 2 showed an analysis based on electronic design for the first implementation. For the digital electronic design, some functions as multiplication, addition and subtraction are simple for digital electronic implementations. Floating-point division and square root are harder to implement due to the complexity of the algorithms. For the division function, if the divided number is $2^m$ (m is integer), it is simple in the digital electronic design, which just shift right or left every bit. But if the divided number is indivisible by 2, it needs much more resources to calculate the same function.

The CIE L*a*b* space has a well uniformity of the perception, and is designed to approximate human vision, but the cube root and the division in the RGB=>L*a*b* transformation requires high precision floating-point calculation which will take too much FPGA resources and memory. The XYZ colour projection has a good parallelism possibility. The major functions of the transformation are multiplication and addition, which are very simple for the VHDL code design. This colour space needs the information of the acquisition equipment, which limits its reusability. The common drawback of the XYZ and L*a*b* colour space transformations is that they both require to know colour coordinates of the reference white. RGB colour space is chosen because it does not have similar drawback. Furthermore, the basic data are integer. In this case, we chose the RGB colour projection as the first step digital electronic implementation.

*Distance Algorithm Analysis*

This analysis is based on the adaptability to digital electronic design. There is no distance algorithm which is suitable for every type of image correlation requirement. If the underlying data distribution is known or can be well modelled, it is possible to find the best distance function that matches the distribution. Several algorithms are implemented in the processing modules. Future user can choose the suitable colour space corresponding to the compared images. 6 distance algorithms are analyzed in Table 3. The original equations of these distance algorithms are presented in the following papers: RMS [19], WRMS [20], GFC [21], $\Delta E_{RGB}$, $\Delta E_{L*a*b*}$ [22], Mv [23].

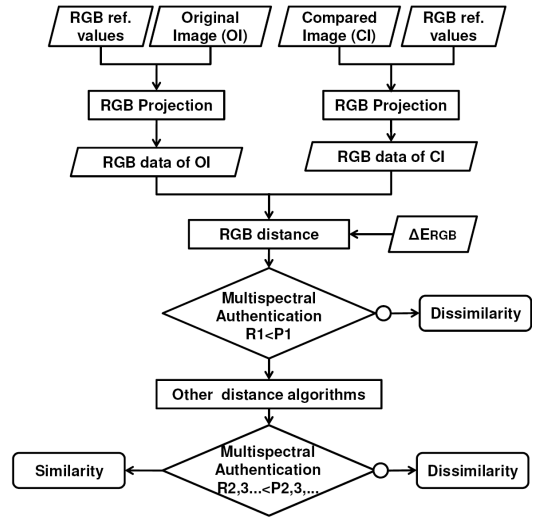

The capacity of the digital design adaptability is as follow: RMS > $\Delta E_{RGB}$ > WRMS > GFC > $\Delta E_{L*a*b*}$ > Mv. RMS algorithm was already implemented in the architecture which was presented in [9]. The authentication process of RGB distance is shown in Fig.5.

*Figure 5. The authentication process using RGB distance algorithm*

## Implementation of RGB Distance Algorithm

The FPGA structure represents an additional challenge for C and SystemC based synthesis tools due to the higher granularity and heterogeneity of FPGA compared to ASIC. The variety of FPGA resources makes the resource selection more difficult for the compiler tools to synthesize high-level C constructs. Several similar resources can be good candidate for one C-construct. The compiler tool has to select the most appropriate resources among all candidate resources.

**Table 2: Comparison of RGB, XYZ and L*a*b***

| Colour space | Advantage | Drawback | Parameter | Colour Family | Principal Function |
|---|---|---|---|---|---|
| RGB | - Data can be coded directly on integer<br>- Additive colour model<br>- CRT visualisation | - Image processing can't be calculated separated. Strong correlation among the colour components. | - Module of human visual system<br>- Materiel dependent | - Primary system CIE standard | - ×<br>- + |
| XYZ | - Luminance and hue components are independent | - Need the information of the acquisition equipments | - Materiel independent | - CIE XYZ standard | - ×<br>- + |
| L*a*b* | - Correspondence between the physical device measurement and the perception of two similar colour<br>- Approximate human vision | - Non-linear transformation<br>- Pertinent only the input data are ≥12-bit (normally 16-bit) RGB colour components<br>- Need the information of the acquisition equipments | - Perceptual system<br>- Human model<br>- Materiel independent | - Perceptually uniform<br>- CIE L*a*b* | - $\sqrt[3]{}$<br>- /<br>- ×<br>- − |

**Table 3: Distance Algorithm Analysis: RMS, WRMS, GFC, $\Delta E_{RGB}$, $\Delta E_{L^*a^*b^*}$ and Mv \***

| Algo. (data type) | Colour projection operation N° | Distance algorithm operation N° | Result type for each step | Parallelism possibility** | Problem definition in electronic design |
|---|---|---|---|---|---|
| RMS (spectra) | 0 | 400 −<br>400 +<br>1 × | Integer<br>Integer<br>Float | Yes<br>No<br>No | - N° of wavelength must be $2^m$ (m=1,2,...9) |
| WRMS (spectra) | 0 | 400 −<br>400 ×<br>400 ×<br>400 +<br>1 ×<br>1 √ | Integer<br>Float<br>Float<br>Float<br>Float<br>Float | Yes<br>Yes<br>Yes<br>No<br>No<br>No | - N° of wavelength must be $2^m$ (m=1,2,...9)<br>- Weights need to be normalized<br>- Square root need more resource |
| GFC (spectra) | 0 | 400 ×<br>400 +<br>400 × 2 ×<br>400 × 2 +<br>400 × 2 √<br>1 × | Integer<br>Integer<br>Integer<br>Integer<br>Float<br>Float | Yes<br>No<br>Yes<br>Yes<br>Yes<br>No | - High precision needs a lot of float number. The result is [0,1] |
| $\Delta E_{RGB}$ (RGB) | 400×3 ×<br>400×3 + | 400 × 3 −<br>400 × 3 ×<br>400 × 2 +<br>400 × 1 √ | Integer<br>Integer<br>Integer<br>Float | Yes<br>Yes<br>Yes<br>Yes | - Square root need more resource |
| $\Delta E_{L^*a^*b^*}$ (L*a*b*) | spectra to XYZ:<br>400×3 ×<br>400×3 +<br>$x_r, y_r, z_r$ values :<br>3 ×<br>3 $\sqrt[3]{}$ | 400 × 3 −<br>400 × 3 ×<br>400 × 2 +<br>400 × 1 √ | Float<br>Float<br>Float<br>Float | Yes<br>Yes<br>Yes<br>Yes | - too complicated for digital design |
| Mv (L*a*b* & spectra) | spectra to XYZ:<br>400×3 ×<br>400×3 +<br>$x_r, y_r, z_r$ values :<br>3 /<br>3 $\sqrt[3]{}$ | 400 −<br>400 × 3 /<br>400 × 3 ×<br>400 × 2 +<br>400 √<br>400 ×<br>400 + | Integer<br>Float<br>Float<br>Float<br>Float<br>Float<br>Float | Yes<br>Yes<br>Yes<br>Yes<br>Yes<br>Yes<br>No | - too complicated for digital design |

**\* Input data are 8-bit integer spectral values obtained by multispectral camera. The spectre field of the multispectral camera is from 380nm to 780nm with 1nm as the minimum unit. For 1 pixel image, the maximum spectre value is 400. The number of the operations is calculated based on this maximum which presents the maximum calculation operations for 1 pixel.**
**\*\* Parallelism possibility means if when we can calculate all the spectra values on parallel at the same time (with ONE cycle clock period on the FPGA board).**

The device utilization summary is based on an FPGA of Xilinx Virtex4 device XC4VLX15. The implementation of RGB distance algorithm contains 2 parts:

– RGB projection: Handel-C [24] with DK Design Suite Tool [25] is used. VHDL version was obtained and synthesized on Xilinx ISE. The result is shown in Table 4.
– $\Delta E_{RGB}$ calculation: VHDL with Xilinx ISE is used. The synthesize result of the most complicated function square root (1st version) is shown in Table 5.

**Table 4: Device Utilization Summary of RGB Projection**

| Device Utilization Summary | | | |
|---|---|---|---|
| Logic Utilization | Used | Available | Utilization |
| N° of Slices | 63 | 6144 | 1% |
| N° of Flip-Flops | 87 | 12288 | 0% |
| N° of LUTs | 84 | 12288 | 0% |

| N° of DSP48s | 4 | 32 | 12% |

**Table 5: Device Utilization Summary of Square Root**

| Device Utilization Summary | | | |
|---|---|---|---|
| Logic Utilization | Used | Available | Utilization |
| N° of Slices | 161 | 6144 | 2% |
| N° of Flip-Flops | 77 | 12288 | 0% |
| N° of LUTs | 292 | 12288 | 2% |
| N° of GCLKs | 1 | 32 | 3% |

The resources for the other reusable units inside the processing module are shown in Table 6.

**Table 6: Resources Utilization Summary of the Reusable Units in the Processing Modules**

| Device Utilization Summary | | |
|---|---|---|
| Reusable units | N° of LUTs | N° of Flip-Flop |
| Communication | 33 | 34 |
| Decode | 12 | 24 |
| Control | 42 | 49 |
| Storage | 48 | 63 |
| Interface | 5 | 4 |

The synthesis frequency of the processing unit which contains RGB distance process is 186MHz. The total logic cells for one RGB distance algorithm processing module is 364. Each unit in the processing module has their own frequencies. The global frequency of the entire processing module is 50MHz.

Table 7 presented the resources utilization summary of the reusable modules in the architecture. The acquisition module is a USB board which is connected to the multispectral camera. More details of the control and storage modules were presented in [4]. For the storage module, extra RAM needs to be used for the multispectral algorithm. The resources in the table just presented the 16M RAM on the FPGA board.

**Table 7: Resources Utilization Summary of the Reusable Module**

| Device Utilization Summary | | |
|---|---|---|
| Reusable module | N° of LUTs | N° of Flip-Flop |
| Control | 278 | 297 |
| Acquisition | 315 | 228 |
| Storage | 280 | 524710 |

Table 8 presented the frequencies of each module. 4 global clocks are used by each type of modules. The network in the system is clockless.

**Table 8: Frequencies Summary of each module**

| Frequency Summary | |
|---|---|
| Module | Frequency (MHz) |
| Control | 150 |
| Acquisition | 77 |
| Storage | 100 |
| Processing | 50 |

**Table9: Resources Utilization Summary of the Entire System**

| Device Utilization Summary | | | |
|---|---|---|---|
| Logic cells | Used | Available | Utilization |
| Entire System | 1237 | 6144 | 20% |

Table 9 presented the global resources of the RGB correlation system with one processing module. The summary showed that only 20% of the slices were used on the FPGA board. More multispectral image correlation algorithms can be implemented using the rest of the resources.

## Conclusion and Perspectives

The advantage of using FPGA include a shorter time to market, ability to re-program in the field to fix bugs, and lower non-recurring engineering costs. Systems can be implemented on an embedded architecture, which need much less energy than computer. The aim of this work was to implement multispectral image correlation in one image analysis adaptive architecture. Several algorithms need to be implemented in the architecture for different kinds of image correlation. This paper presented a part of the RGB distance algorithm implementation. More optimisation and algorithm implementation will be done in the future work.

## References

[1] D. Tzeng, "Spectral-based color separation algorithm development for multiple-ink color reproduction," Ph.D. Dssertation, R.I.T., Rochester, NY, 1999.

[2] E.A. Day, "The Effects of Multi-channel Spectrum Imaging on Perceived Spatial Image Quality and Color Reproduction Accuracy," M.S. Thesis, R.I.T., Rochester, NY, 2003.

[3] V. Vurpillot, A.C. Legrand and A. Trémeau, "Spectral Sensitivity Estimation for Color Camera Calibration," in *IS&T* Proc. The Third European Conference on Colour in Graphics, Imaging, and Vision, CGIV'2006, pp 302-306, Leeds, UK, June 2006.

[4] V. Fresse, A. Aubert and N. Bochard "A Predictive NoC Architecture for Vision Systems Dedicated to Image Analysis," EURASIP Journal on Embedded Systems Volume 2007 Article ID 97929, 13 pages, 2007.

[5] D. Chapiro, "Globally-Asynchronous Locally-Synchronous Systems", PhD thesis, Stanford University, 1984.

[6] J. Muttersbach, T. Villiger, and W. Fichtner, "Practical design of globally-asynchronous locally-synchronous systems," in IEEE Proc. of the Sixth International Symposium on Advanced Research in Asynchronous Circuits and Systems 2000.

[7] D. Galloway, "The transmogrifier C hardware description language and compiler for FPGAs," IEEE Proc. of the Symposium on FPGAs for Custom Computing Machines, pp. 136-144, Napa California, April 1995.

[8] D.C Ku and G.De Micheli, "Hardware C: a language for hardware design," Technical report, Computer System Lab, Stanford University, August 2000.

[9] Z. Larabi, L. Zhang, V. Fresse and A.-C. Legrand, "Rapid Prototyping of Image Analysis Algorithms on an Adaptive FPGA Architecture," in Proc. 15[th] EUSIPCO EURASIP Conf. European Signal Processing, pp. 841-845. Poznan, Poland, 2007.

[10] Xilinx Virtex-4 http://www.xilinx.com/products/silicon_solutions/fpgas/virtex/virtex4/index.htm

[11] B. Fagin and C. Renard, "Field programmable gate arrays and floating point arithmetic," IEEE Transactions on VLSI Systems, vol. 2, pp. 365-367, Sept. 1994.

[12] N. Shirazi, A. Walters, and P. Athanas,"Quantitative analysis of floating point arithmetic on FPGA based custom computing machines," in IEEE Symposium on FPGAs for Custom Computing Machines, pp. 155–162, April 1995.

[13] Z. Luo and M. Martonosi, "Accelerating pipelined integer and floating-point accumulations in configurable hardware with delayed addition techniques," in IEEE Transactions on Computers, vol. 49, pp. 208–218, March 2000.

[14] Y. Dou, S. Vassiliadis, G. K. Kuzmanov, and G. N. Gaydadjiev, "64-bit floating-point FPGA matrix multiplication," in 13th International Symposium on Field-Programmable Gate Arrays, 2005 ACM/SIGNA, pp.86-95, Feb. 2005

[15] P. Hung, H. Fahmy, O. Mencer, and M. J. Flynn, "Fast division algorithm with a small lookup table," in the Proc. of IEEE *Asilomar Conference*, Signals, Systems, and Computers 1999. the 33$^{rd}$ Asilomar Conference, Vol. 2, Issue 1999, pp.1465 – 1468, 1999.

[16] M. D. Ercegovac, T. Lang, J.-M. Muller, and A. Tisserand, "Reciprocation, square root, inverse square root, and some elementary functions using small multipliers," IEEE Transactions on Computers, vol. 2, pp. 628-637, 2000.

[17] D.Strenski, "FPGA Floating Point Performance – a pencil and paper evaluation," Available: http://www.hpcwire.com/hpc/1195762.html

[18] X.Wang, S.Braganza, M.Leeser, "Advanced Components in the Variable Precision Floating-Point Library," in IEEE Proc. of FCCM'06 14$^{th}$ Annual IEEE symposium on Field-Programmable Custom Computing Machines, pp.249-258, April 2006.

[19] F. H. Imai, R. S. Berns and D. Tzeng, "A comparative analysis of spectral reflectance estimated in various spaces using a trichromatic camera system," J. Imaging Sci. Tech. 44, pp. 280-287, 2000.

[20] Y. Zhao, R. Berns, Y. Okumura and L. Taplin, "Improvement of spectral imaging by pigment mapping," *IS&T*, in the Proc. of 13$^{th}$ Color Imaging Conference, pp.40-45, November 2005.

[21] J. Hernández-Andrés, J. Romero, and R. L. Lee, Jr, "Colorimetric and spectroradiometric characteristics of narrow-field-of-view clear skylight in Granada," J. of the Optical Society of America A vol.18, pp. 412-420, 2001.

[22] F. H. Imai, M. R. Rosen and R. S. Berns, "Comparative study of metrics for spectral match quality," in Proc. of the 1$^{st}$ European Conference on Color in Graphics, CGIV'2002, Imaging and Vision, IS&T, Springfield, VA, 2002, pp. 492-496, 2002.

[23] J.A.Stephen Viggiano, "Perception-referenced method for comparison of radiance ratio spectra and its application as an index of metamerism," in the Proc. of SPIE 9$^{th}$ Congress of the International Colour Association, Robert Chung, Allan Rodrigues, Editors, vol.4421, pp. 701-704, June 2002.

[24] Celoxica http://www.celoxica.com/products/dk/default.asp

[25] Handel C http://www.celoxica.com/technology/c_design/handel-c.asp

## Author Biography

*Linlin Zhang received her BS in Optic Electronic Information Engineering from the University of Tianjin (2000). She graduated with M.S in Optic, Photonic and Hyper frequency from Jean Monnet University and French Engineer degree in Optic-electronic from ISTASE (2006). Recently, she is working on her PhD thesis in LaHC CNRS 5516, Saint Etienne, France. Her work has focused on the adaptation of multispectral image analysis algorithms for reconfigurable architecture.*

*Anne-Claire Legrand received her PhD in signal and image processing from Burgundy University, France (2002). She is involved in LIGIV laboratory, University of Saint Etienne, France. Her research interests are focused on spectral imaging and colour science for digital acquisition. She works on design and evaluation of similarity metrics for spectral imaging.*

*Virginie Fresse received her PhD in Electronics and Image Processing from INSA Rennes, France (2001). She is an associate professor in the Hubert Curien Laboratory, University of Saint Etienne, France. Her research interests are adaptive NoC-based FPGA architectures, High level synthesis, implementations of image algorithms on real-time embedded systems.*